\documentclass[a4paper]{article}

\usepackage{INTERSPEECH2018}
\usepackage{flushend}
\usepackage[section]{placeins}

\usepackage{amsmath,amssymb}
\DeclareMathOperator{\E}{\mathbb{E}}
\usepackage[most]{tcolorbox}
\usepackage{xcolor}

\usepackage{mathrsfs}
\newcommand{\ve}[1]{\textbf{#1}}		

\title{Whispered-to-voiced Alaryngeal Speech Conversion with Generative Adversarial Networks}

\name{Santiago Pascual$^1$, Antonio Bonafonte$^1$, Joan Serr\`a$^2$, Jose A.~Gonzalez$^3$}
\address{
  $^1$Universitat Polit\`ecnica de Catalunya, Barcelona, Spain\\
  $^2$Telef\'onica Research, Barcelona, Spain\\
  $^3$Universidad de M\'alaga, M\'alaga, Spain}
\email{santi.pascual@upc.edu}

\begin{document}

\maketitle

\begin{abstract}
Most methods of voice restoration for patients suffering from aphonia either produce whispered or monotone speech. Apart from intelligibility, this type of speech lacks expressiveness and naturalness due to the absence of pitch (whispered speech) or artificial generation of it (monotone speech). Existing techniques to restore prosodic information typically combine a vocoder, which parameterises the speech signal, with machine learning techniques that predict prosodic information. In contrast, this paper describes an end-to-end neural approach for estimating a fully-voiced speech waveform from whispered alaryngeal speech. By adapting our previous work in speech enhancement with generative adversarial networks, we develop a speaker-dependent model to perform whispered-to-voiced speech conversion. Preliminary qualitative results show effectiveness in re-generating voiced speech, with the creation of realistic pitch contours. 

\end{abstract}

\vspace{0.2cm}
\noindent\textbf{Index Terms}: pitch restoration, whispered speech, generative adversarial networks, alaryngeal speech.

\section{Introduction}
\label{sec:intro}

Whispered speech refers to a form of spoken communication in which the vocal folds do not vibrate and, therefore, there is no periodic glottal excitation. This can be intentional (e.g.,~speaking in whispers), or as a result of disease or trauma (e.g.,~patients suffering from aphonia after a total laryngectomy). The lack of pitch reduces the expressiveness and naturalness of the voice. Moreover, it can be a serious impediment for speech intelligibility in tonal languages~\cite{Chen2018} or in the presence of other interfering sources (i.e.,~cocktail party problem~\cite{Popham2018}). The conversion from whispered to voiced speech, either by reconstructing partially existent pitch contours or by generating completely new ones, is an area of research that not only has relevant practical and real-world applications, but also fosters the development of advanced speech conversion systems. 

In general, existing methods 
for whispered-to-voiced speech conversion either follow a data-driven or an analysis-by-synthesis approach. In the data-driven approach, machine learning is used to estimate the pitch from the available speech parameters (e.g.,~mel frequency cepstral coefficients; MFCCs). Then, a vocoder is used to synthesize speech from those by, for instance, predicting fundamental frequencies and voiced/unvoiced decisions from frame-based spectral information of the whispered signal, using Gaussian mixture models (GMMs)~\cite{Toda2008,Nakamura2011,Nakamura2012} or deep neural networks~\cite{Gonzalez2017a}. The analysis-by-synthesis approach follows a similar methodology to code-excited linear prediction~\cite{Morris2002,Ahmadi2008,Sharifzadeh2010}. To estimate pitch parameters, a common strategy is to derive those from other parameters available in the whispered signal, such as estimated speech formants~\cite{Li2014}.

A key application of whisper-to-voiced speech conversion is to provide individuals with aphonia with a more naturally sounding voice. People who have their larynx removed as a treatment for cancer inevitably lose their voice. To speak again, laryngectomees can resort to a number of methods, such as the voice valve, which produces an unnatural, whispered sound, or the electrolarynx, a vibrating device placed against the neck that generates the lost glottal excitation but, nonetheless, produces a robotic voice due to its constant vibration. In recent years, the use of whisper-to-speech reconstruction methods~\cite{Toda2008,Nakamura2011,Nakamura2012,Fuchs2012}, or silent speech interfaces~\cite{Denby2010,Gonzalez2017a,Gonzalez2017b} in which an acoustic signal is synthesised from non-audible speech-related biosignals such as the movements of the speech organs, have started to be investigated to provide laryngectomees with a better and more naturally sounding voice.

In this paper, and in contrast to previous approaches, we present a speaker-dependent end-to-end model for voiced speech generation based on generative adversarial networks (GANs)~\cite{GANsPaper}. With an end-to-end model directly performing the conversion between waveforms, we avoid the explicit extraction of spectral information, the error-prone prediction of intermediate parameters like pitch, and the use of a vocoder to synthesize speech from such intermediate parameters. With a GAN learning the mapping from whispered to voiced speech, we avoid the specification of an error loss over raw audio and make our model truly generative, thus being able to produce new realistic pitch contours. Our results show this novel pitch generation as an implicit process of waveform restoration. To evaluate our proposal, we compared the pitch contour distributions predicted by our proposal with those obtained by a regression model, observing that our proposal is able to attain more natural pitch contours than those predicted by a regression model, including a more realistic variance factor that relates to more expressiveness.

The remainder of the paper is structured as follows. In section~\ref{sec:method} we describe the method used to restore the voiced speech with GANs. The experimental setup is described in section~\ref{sec:setup}, which includes descriptions of the dataset, an RNN baseline and the hyper-parameters for our GAN model. Finally, sections~\ref{sec:results} and~\ref{sec:conclusions} contain the discussions of results and conclusions respectively.


\section{Generative Adversarial Networks for Voiced Speech Restoration}
\label{sec:method}

The proposed model is an improvement over our previous work on speech enhancement using GANs (SEGAN)~\cite{pascual2017segan,pascual2018koreansegan} in order to handle speech reconstruction tasks. SEGAN was designed as a speaker- and noise-agnostic model to generate clean/enhanced versions of aligned noisy speech signals. From now on, we change signal names for the new task, so we rather work with natural, voiced (i.e. restored) and whispered speech signals. To adapt the architecture to the task of voiced speech restoration, we decide to remove the audio alignment requirement, as the data we use has slight misalignments between input and output speech (see Section \ref{ssec:task} for more details). In addition, we introduce a number of improvements that consistently stabilize and facilitate its training after direct regularization over the waveform is removed. These modifications also refine the generated quality at the generator output when regression is removed.

\subsection{SEGAN}

We now outline the most basic aspects of SEGAN, specifically highlighting the ones that have been subject to change. For the sake of brevity we refer the reader to the original paper and code~\cite{pascual2017segan} for more detailed explanations on the old architecture and setup. The SEGAN generator network ($G$) embeds an input noisy waveform chunk into the latent space via a convolutional encoder. Then, the reconstruction is made in the decoder by `deconvolving' back the latent signals into the time domain. $G$ features skip connections with constant factors (acting as identity functions) to promote that low-level features could escape a potentially unnecessary compression from the encoder. Such skip connections also improve training stability, as they allow gradients to flow better across the deep structure of G, which has a total of 22~layers. In the denoising setup, an L$_1$ regularization term helped centering output predictions around 0, discouraging $G$ to explore bizarre amplitude magnitudes that could make the discriminator network ($D$) converge to easy discriminative solutions for the fake adversarial case.

\subsection{Adapted SEGAN} 

The SEGAN architecture has been adapted to cope with misalignments in the input/output signals as mentioned before, as well as to achieve a more stable architecture and to produce better quality outputs. In the current setup, similarly to the original SEGAN mechanism, we inject whisper data to $G$, which compresses it and then recovers a version of the utterance with prosodic information. 
To cope with misalignments, we get rid of the L$_1$ regularization term, as this was forcing a one-to-one correspondence between audio samples, assuming input and output had the same phase. In its place we use a softer regularization which works in the spectral domain, similar to the one used in the parallel Wavenet~\cite{oord2017parallel}. We use a non-averaged version of this loss though, as we work with large frames during training (16,384 samples per sequence), and averaging the spectral frames over this large span could be ineffective. Moreover, we calculate the loss as an absolute distance in decibels between the generated speech and the natural one.

\begin{figure}[!t]
\centering
\includegraphics[width=0.85\linewidth]{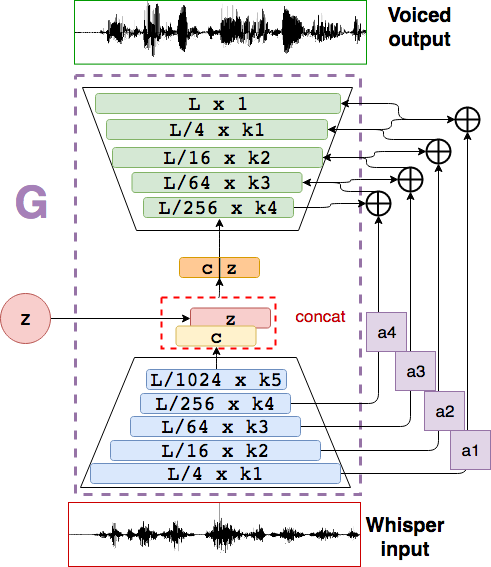}
\caption{Generator network architecture. Skip connection with learnable $\ve{a}_l$ are depicted with purple boxes. These are summed to each intermediate activation of the decoder. Encoder and decoder are like in the original SEGAN~\cite{pascual2017segan}, but with half the amount of layers and doubled pooling per layer.}
\label{fig:silentsegan}
\end{figure}

The spectral regularization is added to the adversarial loss coming from $D$ with a weighting factor $\lambda$. In SEGAN, $D$ is a learnable comparative loss function between natural or voiced signals and whispered ones. This means we have a \texttt{(natural, whispered)} paired input as a real batch sample and \texttt{(voiced, whispered)} as a fake batch sample. In contrast, $G$ has to make \texttt{(voiced,whispered)} true, thus being the adversarial objective. In the current setup, we add an additional fake signal in $D$ that will enforce the preservation of intelligibility when we forward data through $G$: the \texttt{(natural,random\_natural\_shuffle)} pair. This pair tries to send messages to $G$ about a bad behavior whenever the content between both chunks, the one coming from $G$ and the reference one, changes. Note that we are using the least-squares GAN form (LSGAN) in the adversarial component, so our loss functions, for $D$ and $G$ respectively, become
\begin{equation*}
\begin{aligned}
\underset{D}\min~V(D) & = \frac{1}{3}\E_{\ve{x},\tilde{\ve{w}}\sim p_{\text{data}}(\ve{x}, \tilde{\ve{w}})}[(D(\ve{x},\tilde{\ve{w}}) - 1)^{2}] +\\
   & + \frac{1}{3}\E_{\ve{z}\sim p_{\ve{z}}(\ve{z}),\tilde{\ve{w}}\sim p_{\text{data}}(\tilde{\ve{w}})}[D(G(\ve{z},\tilde{\ve{w}}),\tilde{\ve{w}})^{2}] \\
   & + \frac{1}{3}\E_{\ve{x},\ve{x}^{r}\sim p_{\text{data}}(\ve{x})}[D(\ve{x},\ve{x}^r)^{2}] \\
\underset{G}\min~V(G) & = \E_{\ve{z}\sim p_{\ve{z}}(\ve{z}),\tilde{\ve{w}}\sim p_{\text{data}}(\tilde{\ve{w}})}[(D(G(\ve{z},\tilde{\ve{w}}),\tilde{\ve{w}}) - 1)^{2}],
\end{aligned}
\end{equation*}

where $\tilde{\ve{w}}\in\mathbb{R}^T$ is the whispered utterance, $\ve{x}\in\mathbb{R}^T$ is the natural speech, $\ve{x}^r\in\mathbb{R}^T$ is a randomly chosen natural chunk within the batch, $G(\ve{z}, \tilde{\ve{w}})\in\mathbb{R}^T$ is the enhanced speech, and [$D(\ve{x}, \tilde{\ve{w}}), D(G(\ve{z}, \tilde{\ve{w}}), \tilde{\ve{w}}), D(\ve{x}, \ve{x}^r)$] are the discriminator decisions for each input pair. All of these signals are vectors of length $T$ samples except for $D$ outputs, which are scalars. $T$ is a hyper-parameter fixed during training but it is variable during test inference.

After removing the regularization factor L$_1$, the generator output can explore large amplitudes whilst adapting to mimic the speech distribution. As a matter of fact, this collapsed the training whenever the $\tanh$ activation was placed in the output layer of $G$ to bound its output to $[-1, 1]$, because the amplitude grew quickly with aggressive gradient updates and $\tanh$ would not allow $G$ to properly update anymore due to saturation. The way to correct this was bounding the gradient of $D$ by applying spectral normalization as proposed in~\cite{miyato2018spectral}. The discriminator does not have any batch normalization technique in this implementation, and its architecture is the same as in our previous work.

\begin{figure*}[htpb!]
\centering
\fbox{\includegraphics[width=0.25\linewidth, height=130pt]{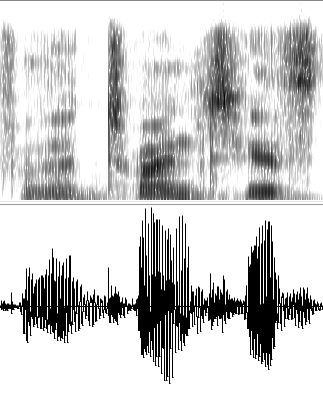}}
\fbox{\includegraphics[width=0.25\linewidth, height=130pt]{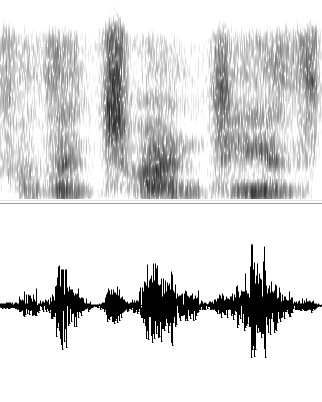}}
\fbox{\includegraphics[width=0.25\linewidth, height=130pt]{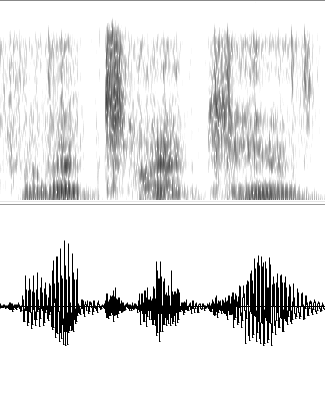}}
\caption{\label{fig:wavs_specs}From left to right: natural speech, whispered speech (input to $G$) and output from $G$ as voiced signal.}
\end{figure*}

The new design of $G$ is shown in Figure~\ref{fig:silentsegan}. It remains as a fully convolutional encoder-decoder structure with skip connections, but with two changes. First, we reduce the number of layers by augmenting the pooling factor from 2 to 4 at every encoder-decoder layer. This is in line with preliminary experiments on the denoising task, where increasing pooling has been effective to improve objective scores for that task. Second, we introduce learnable skip connections, and these are now summed instead of concatenated to decoder feature maps. We thus have now learnable vectors $\ve{a}_{l}$ which multiply every channel of its corresponding shuttle layer $l$ by a scalar factor $\alpha_{l,k}$. These factors are all initialized to one. Hence, at the $j$-th decoder layer input we have the addition of the $l$-th encoder layer response following 
\begin{equation*}
 \ve{h}_{j} = \ve{h}_{j-1} + \ve{a}_l\odot\ve{h}_{l} ,
\end{equation*}
where $\odot$ is an element-wise product along channels.

\section{Experimental Setup}
\label{sec:setup}

To evaluate the performance of our technique, a clinical application involving the generation of audible speech from captured movement of the speech articulators is tested. More details about the experimental setup in terms of dataset, baseline and hyper-parameters for our proposed approach are given below.

\subsection{Task and Dataset}
\label{ssec:task}

In our previous work~\cite{Gonzalez2017a,Gonzalez2017b}, a silent speech system aimed at helping laryngectomy patients to recover their voices was described. The system comprised an articulator motion capture device~\cite{Fagan2008}, which monitored the movement of the lips and tongue by tracking the magnetic field generated by small magnets attached to them, and a synthesis module, which generated speech from articulatory data. To generate speech acoustics, recurrent neural networks (RNNs) trained on parallel articulatory and speech data were used. The speech produced by this system had a reasonable quality when evaluated on normal speakers, but it was not completely natural owing to limitations when estimating the pitch (i.e.,~the capturing device did not have access to any information about the glottal excitation). 

In this work, we are interested on determining whether the proposed adapted SEGAN could improve those signals by generating more natural and realistic prosodic contours. To evaluate this, we have articulatory and speech data available, recorded simultaneously for 6 healthy British subjects (2 females and 4 males). Each speaker has recorded a random subset of the CMU Arctic corpus~\cite{Arctic} (25 minutes for each speaker, approximately). Then, whispered speech was generated from the articulatory data by using the RNN-based articulatory-to-speech synthesiser described in~\cite{Gonzalez2017b}. In this work, these whispered signals are taken as the input to SEGAN, which acts as a post-filter enhancing the naturalness of the signals. For each whispered signal we have a natural version, which is the original speech signal recorded by the subject. To simplify our first modeling approach we used one male and one female speakers, namely M4 and F1, and built two speaker-dependent SEGAN models. These speakers are selected for the better level of intelligibility of their whisper data within their genders. We want to note, however, that both female speakers are less intelligible in their whisper form than male speakers. These two speakers' data is split into two sets: (1) training, with approximately 90\% of the utterances and (2) test, with the remaining approximate 10\%. In order to have augmented data we follow the same chunking method as in our previous work~\cite{pascual2017segan} but window strides are one order of magnitude smaller. Hence we have a canvas of 16,384 samples ($\approx$ 1 second at 16kHz) every 50\,ms, in contrast with the previous 0.5\,s. 

\begin{figure*}[!t]
\minipage{0.5\textwidth}
\includegraphics[width=\linewidth]{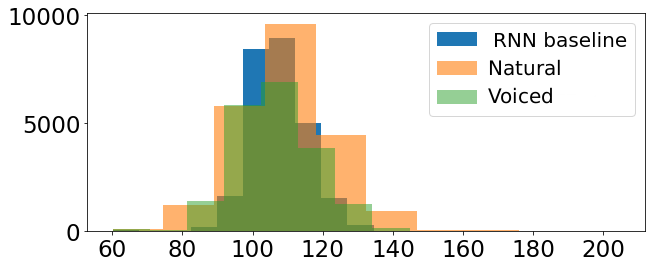}
\endminipage\hfill
\minipage{0.5\textwidth}
\includegraphics[width=\linewidth]{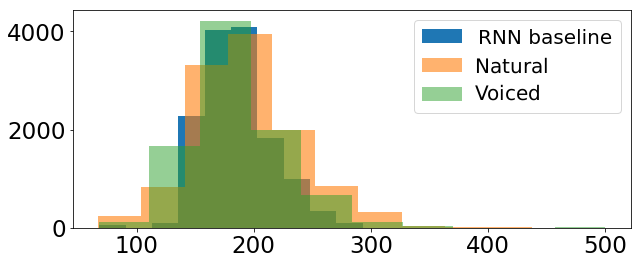}
\endminipage
\caption{\label{fig:meanpitch_hists} Histograms of pitch values in Hertz per utterance for male speaker (left) and female speaker (right). The three systems appearing are natural signals; RNN baseline voiced predictions with vocoder features; and voiced speech using SEGAN.}
\end{figure*}

\begin{figure*}[!t]
\includegraphics[width=\linewidth]{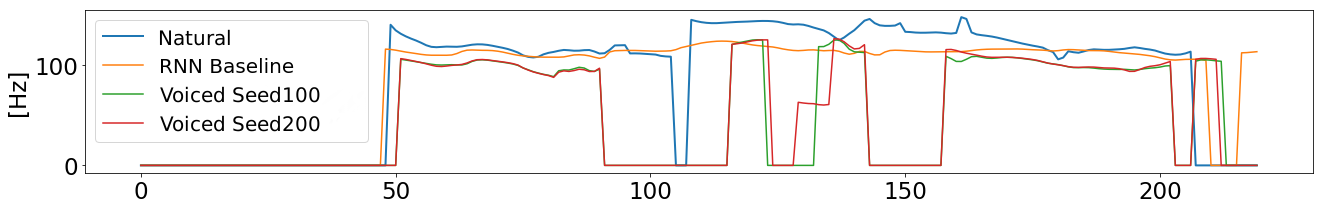}
\caption{\label{fig:pitch_contours} Section of pitch contour of a test utterance of the male speaker calculated with Ahocoder from 4 different sources: Natural data (blue); RNN baseline (orange); Voiced with seed 100 (green) and voiced with seed 200 (red). Here it is shown how changing the seed indeed creates different plausible contours.}
\end{figure*}

\subsection{SEGAN Setup}

We use the same kernel widths of 31 as we had in~\cite{pascual2017segan}, both when encoding and decoding and for both $G$ and $D$ networks. The feature maps are incremental in the encoder and decremental in the decoder, having \{64, 128, 256, 512, 1024, 512, 256, 128, 64, 1\} in the generator and \{64, 128, 256, 512, 1024\} in the discriminator convolutional structures. The discriminator has a linear layer at the end with a single output neuron, as in the original SEGAN setup. The latent space is constructed with the concatenation of the thought vector $\ve{c}\in\mathbb{R}^{\frac{T}{1024}\times 1024}$ with the noise vector $\ve{z}\in\mathbb{R}^{\frac{T}{1024}\times 1024}$, where $\ve{z}\sim \mathcal{N}(0,I)$.
Both networks are trained with Adam~\cite{kingma2014adam} optimizer, with the two-timescale update rule (TTUR)~\cite{heusel2017gans}, such that $D$ will have a four times faster learning rate to virtually emulate many iterations in $D$ prior to updating G. This way, we have $D$ learning rate 0.0004 and $G$ learning rate 0.0001, with $\beta_1 = 0$ and $\beta_2 = 0.9$, which are the same schedules based on recent successful approaches to faster and stable convergent adversarial training~\cite{zhang2018self}. All signals processed by the GAN, either in the input/output of $G$ or the input of D, are pre-emphasized with a 0.95 factor, as it proved to help coping with some high-frequency artifacts in the de-noising setup. When we generate voiced data out of $G$ we de-emphasize it with the same factor to get the final result.

\subsection{Baseline}

To assess the performance of SEGAN in this task we have as reference the RNN-based articulatory-to-speech system from our previous work~\cite{Gonzalez2017b} and the natural data for each modeled speaker. The recurrent model is used to predict both the spectral (i.e.,~MFCCs) and pitch parameters (i.e.,~fundamental frequency, aperiodicities and unvoiced-voiced decision) from the articulatory data, so the source is articulatory data and not whispered speech in that case. The STRAIGHT vocoder~\cite{Kawahara1999} is then employed to synthesise the waveform from the predicted parameters.

\section{Results}
\label{sec:results}

We analyze the statistics of the generated pitch contours for the RNN, SEGAN and natural data. Figure~\ref{fig:meanpitch_hists} depicts the histograms of all contours extracted from predicted/natural waveforms. Ahocoder~\cite{ahocoder2014} was used to extract $\log$F0 curves, which are then converted to Hertz scale. Then, all voiced frames were selected and concatenated per each of the three systems. We come up with a long stream for each system and for the two genders. It can seen that, for both genders, voiced histograms (corresponding to SEGAN) have a broader variance than RNN ones, closer to the natural signal shape. This is understandable if we consider that the RNN was trained with a regression criterion that optimizes its output towards the mean of the pitch distribution. This ends up producing a monotonic prosody effect, normally manifested as a robotic sounding that can be heard in the audio samples referenced below. This indicates that the adversarial procedure can generate more natural pitch values.

Figure~\ref{fig:pitch_contours} shows pitch contours generated by SEGAN with different random seeds. We have to note that each random seed generates a different latent vector $\ve{z}$, so the stochasticity creates novel curves that look plausible. It also can be noted that SEGAN made some errors in determining the correct voicing decision for some speech segments. We may enforce a better behavior in a future version of the system with an auxiliary unvoiced/voiced classifier in the output of G.

Finally, figure~\ref{fig:wavs_specs} shows examples of waveforms and spectrograms for natural, whispered and voiced signals. We can appreciate how, for a small chunk of waveform, the generator network is able to refine low frequencies and gets rid of high frequency noises to approximate the natural data. Preliminary listening tests suggest that this model can achieve a good natural voiced version of the speech, but some artifacts intrinsic to the convolutional architecture (specially in high-frequencies) have to be palliated. This observation is in line with what was also prompted in the WaveGAN~\cite{donahue2018synthesizing} work, and this is also one of the potential reasons of the effectiveness of using pre-emphasis. We refer the reader to the audio samples to have a feeling of the current quality of our system~\footnote{http://veu.talp.cat/whispersegan/}.



\section{Conclusions}
\label{sec:conclusions}

We presented a speaker-dependent end-to-end generative adversarial network to act as a post-filter of whispered speech to deal with a pathological application. We adapted our previous speech enhancement GAN architecture to overcome misalignment issues and still obtained a stable GAN architecture to reconstruct voiced speech. The model is able to generate novel pitch contours by only seeing the whispered version of the speech at its input. The method generates richer curves than the baseline, which sounds monotonic in terms of prosody. Future lines of work include an even more end-to-end approach by going sensor-to-speech. Also, further study is required to alleviate intrinsic high frequency artifacts provoked by the type of decimation-interpolation architecture we base our design on.

\section{Acknowledgements}
This research was supported by the project TEC2015-69266-P (MINECO/FEDER, UE).

\clearpage

\bibliographystyle{IEEEtran}

\bibliography{mybib}


\end{document}